\documentstyle[12pt]{article}

\begin{document}

\baselineskip=18.6pt plus 0.2pt minus 0.1pt


\makeatletter
\@addtoreset{equation}{section}
\renewcommand{\theequation}{\thesection.\arabic{equation}}
\begin{titlepage}
\title{
\hfill\parbox{4cm} {\normalsize UFR-HEP/02-01\\hep-th/0201155}\\
\vspace{1cm}
     {\bf       Manifolds of
      $G_2$  Holonomy   from  $ N=4$ Sigma Model  }
}
\author{
Adil Belhaj\thanks{{\tt ufrhep@fsr.ac.ma}}
  {}
\\[7pt]
{\it  High Energy Physics Laboratory, Faculty of sciences, Rabat,
Morocco} }

\maketitle \thispagestyle{empty}
\begin{abstract}
Using  two dimensional ($2D$)  $N=4$  sigma model, with $U(1)^r$ gauge symmetry,
  and introducing the
ADE Cartan matrices as gauge matrix charges, we build " toric"
hyper-Kahler eight real dimensional  manifolds $X_8$. Dividing by one  toric geometry
circle action
of $X_8$ manifolds, we present  examples describing   quotients $X_7={X_8\over U(1)}$
of $G_2$ holonomy. In particular,
for the $ A_r$ Cartan
 matrix,
  the quotient space is a cone on  a $ {S^2}$ bundle over
     $r$ intersecting   $\bf WCP^2_{(1,2,1)}$ weighted  projective spaces according
      to the $A_r$
      Dynkin diagram.

\end{abstract}
\newpage
\newpage
\end{titlepage}
\newpage
\def\be{\begin{equation}}
\def\ee{\end{equation}}
\def\bea{\begin{eqnarray}}
\def\eea{\end{eqnarray}}
\def\nn{\nonumber}
\def\l{\lambda}
\def\t{\times}
\def\[{\bigl[}
\def\]{\bigr]}
\def\({\bigl(}
\def\){\bigr)}
\def\p{\partial}
\def\o{\over}
\def\ta{\tau}
\def\cm{\cal M}
\def\R{\bf R}
\def\b{\beta}
\def\a{\alpha}
\section{Introduction}
  Over the few past years, there has been an increasing interest in  studying string
  dualities.   One of the important consequence of  these studies  is that all
 superstring models are equivalent in the sense that they
 correspond to different limits in moduli space of the same
 theory, called M-theory  \cite{1,2,3}. The latter,  which is considered nowadays as the best candidate
 for the unification of the weak and strong coupling sectors of
  superstring models,
   is described,
  at low energies, by an eleven dimensional
 supergravity theory.\par
  More recently, a special interest has been given to the
  compactification of the M-theory on seven real manifolds $X_7$ with non trivial
  holonomy. This interest is due  to the fact that  these manifolds provide a
  potential point of contact with low energy semi realistic
  physics from M-theory. In particular, one can obtain four
  dimensional theory with $N=1$ supersymmetry by compactifying
  M-theory     on $R^{1,3}\t X_7$ where $X_7$ a seven manifold
  with   $ G_2$ holonomy [4-11]\footnote{ The latter is the maximal subgroup of $SO(7)$
  which can break the eight dimensional spinoriel representation  of  $SO(7)$ to the seven
  fundamental representation of $ G_2$ plus  one  singlet ( $ 8\to 7+1)$.}.
In this regard, the  $N=1$ four dimensional resulting physics
 models depend on geometric properties of  $X_7$. For instance, if
 $X_7$ is smooth, the low energy theory contains, in addition
  to
 $N=1$
 supergravity, only abelian gauge group and neutral chiral
 multiplets. However,
  non abelian  gauge symmetries with chiral fermions   can be  obtained
  by considering limits where $X_7$ develops singularities [10,11].
  For this reason, it is interesting to study  M-theory on singular
  seven  manifold with $G_2$ holonomy.
  Following \cite{11}, an interesting  analysis for
 building such spaces  is to consider the
 quotient of  conical hyper-Kahler eight manifolds $X_8$ by a
 $U(1)$ symmetry. This approach, which is called the
 unfolding of the singularity, guarantees the $G_2$ holonomy
 group
 of the quotient space
    $X_8\over U(1)$. A remarkable feature of this method,
    which may be   related to  two dimensional  $(2D)$ $N=4$  sigma model
     Calabi-Yau
    fourfolds construction $CY^4$, is that  the $X_8\over U(1)$ space solutions
     differ by
    what kind of $U(1)$
   symmetry is chosen and moreover the matter fields, in four dimensions,
    are obtained using  the  techniques of the geometric engineering of quantum field
    theory [12-15].\par
    The aim of this work is to contribute in this direction by
    considering models with $N=4$ $2D$ sigma model ADE Cartan
    matrix gauge  charges for  building $X_8$ manifolds. This study is  motivated by
    the following points:
     (1)  Actually, these vector charges
     go beyond of the one given in   the first
    example studied in [11], where the matrix charge of the
    hypermultiplets
     $\phi_i$, under $U(1)^r$ gauge symmetry,  was
    \be q^a_i=-\delta^a_{i-1}+\delta^a_{i},\quad a=1,\ldots,r.
    \ee
(2)  They  may give an analogue connection  appearing between the
toric Calabi-Yau geometries, used in string theory
compactifications, and the structures  of ADE Lie algebras. The
latter  may lead to a similar analysis of the geometric
engineering method of quantum field theory embedded in string
theory.
\\ It was suggested in [11] that the unfolding of the singularity may be
   adapted to  others  examples   of $ X_8$ manifolds, in particular
 toric-hyper-Kahler manifolds.  In this paper,  we  would like to present
 a  new class  of these  $ X_8$ manifolds, which  will be  called
  toric-hyper-Kahler eight manifods $X_8$, with  the Calabi-Yau
  condition in sigma model construction given by
  \be \sum_i q^a_i=0.
    \ee
We will refer to such manifolds  as Calabi-Yau fourfolds.   Then
we give their  quotients by a $U(1)$ group symmetry using toric
geometry circle actions. Our way involves two steps:
\\  First  we introduce  the ADE Cartan matrices as $2D$ $N=4$ $U(1)^{r}$
linear sigma model matrix gauge charges.  Second mimicking the
analysis of \cite{11} and
 using toric
geometry circle actions,  we discuss the  construction of  a new
class of the quotients $X_8\over U(1)$ of  $G_2$ holonomy group. In particular, for
the $A_r$ Cartan matrix,
  the quotient space is a cone on a  $ {S^2}$  bundle over
     $r$ intersecting   $\bf WCP^2_{(1,2,1)}$  weighted projective spaces according to the $A_r$
     Dynkin geometry.\\
    The organization of this paper is as follows: In section 2, we
give an overview  on  aspects  of $2D$ $N=4$   linear sigma model.
Then we give  examples illustrating  the field theoretical construction of hyper-Kahler
 manifolds. In section 3, we
introduce the ADE Cartan matrices as  matrix gauge charges in  the
$2D$ $N=4$ field theory construction of $X_8$ manifolds. For the
$A_r$ Lie algebra, the moduli space of the classical theory is
given by the cotangent bundle over  $r$ intersecting $\bf
WPC^2_{1,2,1}$ weighted projective spaces according to the $A_r$
Dynkin graph, extending the $A_r$ singularity of K3 surfaces
described by $ N=2$
 type IIA superstring  sigma model used in the geometric engineering method. In section 4, we
identify  the  $U(1)$ symmetry group with   the  toric geometry
circle actions  of $X_8$ to   present  quotients  $X_7= {X_8\over
U(1)}$ of $G_2$ holonomy.  For the $A_r$ Cartan matrix gauge
charge,
  the geometry  is a cone on a  $ {S^2}$ bundle over
     $r$ intersecting   $\bf WCP^2_{(1,2,1)}$  weighted projective spaces according to the $A_r$
     Dynkin diagram.  Discussion and conclusion will be given in section 5.

    \section{$N=4$ sigma model approach}
    In this section we review the main lines of the $N=4$ sigma model
    approach  for building the hyper-Kahler manifolds involved in the study
    of superstring, M-theory  and F-theory, compactifications, Yang Mills small instantan
     singularities
     and more general
     in supersymmetric field theories with eight supercharges [16,17,18].
     For this purpose, consider
          $2D$  $N=4$ supersymmetric $U(1)^r$ gauge theory  with   $n$
          hypermultiplets
        $\phi_i$
        $( i=1,\ldots,n) $
         of a  matrix  charge
         $q_i^a$  $ (a=1,\ldots,r)$, under $U(1)^r$ gauge symmetry,
         and $r$ 3-vectors FI coupling $\vec \xi_a$. The equations
         defining the hyper-Kahler
         moduli space  of this classical gauge theory  are given by
         the following D-terms
         \be
         \sum\limits_i q^i_a\[\phi_i^{\alpha}{\bar \phi}_{i \beta}
         +\phi_{i \beta}
        {\bar \phi}_i^{\alpha}\]=\vec \xi_a \vec\sigma^\alpha_\beta.
        \ee
      The double index $(i,{\alpha})$ of
          $\phi_i^{\alpha}$'s
        refer to    component field doublets of the $n$
        hypermultiplets,  and  $\vec \sigma^\alpha_\beta$ are the
        traceless $2\t2$ Pauli matrices.
        For later use it is interesting to note the following
        points:\\
       (1) Equations (2.1)
        have a formal analogy  with the   D-flatness equations
         of $2D$  $N=2$  $U(1)^r$  toric  sigma model involved in the
         study of  type II superstring compactifications on ALE spaces
         with ADE singularities[13]. The latters are given  by:
\be \sum \limits _{i=1}^{n} q_i^a |x_i|^2=R_a,\quad a=1,\ldots,r
\ee where $r$ is the rank of the ADE Lie algebras and  where $q_i^a$,
up some details, the minus of the corresponding Cartan  matrices
satisfying  the Calabi-Yau condition \be \sum \limits _{i=1}^{n}
q_i^a =0.
 \ee
Equation (2.2)  has a nice geometrical interpretation in terms of
toric geometry. This  has been a  beautiful  interplay
      between  $2D$ $ N=2$ sigma models and toric geometry.   In this way,
     (2.2) have a  toric diagram   which
      consists of   $n$ vertices  $\{{v_i}\}$,  in  the standard lattice
       $ \bf Z^{n-r}$,  satisfying the following
       constraint
       equations
 \be
\sum \limits _{i=1}^{n} q_i^a v_ i=0, \quad  a=1,\ldots,r. \ee For
instance,  if one takes $r=1$,  this space describes the  $n-1$
dimensional weighted  projective  with weights   $ q_i$.\\ (2) For
each $U(1)$ factor, there are three real constraint equations
transforming as  an iso-triplet of $SU(2)$ R-symmetry ($SU(2)_R$)
acting on the hyper-Kahler structures.
\\(3) Using the $SU(2)_R$ transformations \bea
\phi^{\alpha}&=&\varepsilon^{\alpha\beta}\phi_ {\beta},\quad
\varepsilon_{12}=\varepsilon^{21}=1\\
\overline{(\phi^\alpha)}&=&\overline{\phi}_\alpha,\nn
 \eea and
replacing the Pauli matrices by their expressions, the identities
(2.1) can be
 split as
follows: \bea \sum\limits_{i=1}^n q_i^a(
|\phi^1_i|^2-|\phi^2_i|^2) &= &\xi^3 _a\\ \sum\limits_{i=1}^nq_i^a
\phi^1_i \overline{\phi}_{i}^2&=&\xi^1_a+i{\xi^2}_a
\\
\sum\limits_{i=1}^n q_i^a\phi^2_i
\overline{\phi}_{i}^1&=&\xi^1_a-i{\xi^2}_a. \eea
Note that these equations have similar features of the description of [16] leaving
 only half the supersymmetry of the gauge model.\\
  (4) After dividing the moduli space
of zero energy states of the classical gauge theory  (2.1)   by
the action of the $U(1)^r$ gauge symmetry,  we find  precisely  a
toric-hyper-Kahler variety $X_{4(k-r)}$ of  $4(k-r)$ real
 dimensions.  This construction is called the hyper-Kahler
 quotient extending the Kahler one involved in $2D$ $N=2$
 toric sigma model [16,17,18].\\
 (5) The solutions of eqs (2.1) depend on the values of the FI
 couplings.  For the case where  $\xi^1=\xi^2=0$  and
$\xi^3 >0$, it is not difficult to see that eqs (2.1) describe the
cotangent bundle over a  toric variety defined by  \be
\sum\limits_{i=1}^n q_i^a|\phi^1_i|^2 = \xi^3 _a.\ee Indeed, if we
set all $\phi^2_i=0$, the $\phi^1_i$'s, modulo the complexified
$U(1)^r$ gauge group, determine a toric variety $ C^n\over
{C^*}^r$ of $2(n-r)$ real dimensions, see equations (2.2). The
equations (2.6-7) mean that the $\phi^2_i$'s define the cotangent
fiber directions over the toric variety given by (2.9).  To see
this feature, we assume that  $\xi^1_a=\xi^2_a=0$, $a=1$  and  we
set $q_i^a=q_i=1$, so we have   \be \sum\limits_{i=1}^n(
|\phi^1_i|^2-|\phi^2_i|^2) = \xi^3 \ee  and \be
\sum\limits_{i=1}^n \phi^1_i \overline{\phi}_{i}^2=0.\ee  Equation
(2.10), for $\phi^2_i=0$,    defines  the  $ \bf CP^{n-1}$
projective space while equation (2.11) means that ${\phi}_{i}^2$
parameterizes the cotangent directions over it. In what follows,
we give two
 extra examples illustrating this analysis and reconsidering the example
given in [11]. In the first example, we consider  a $2D$ $N=4$
$U(1)$ linear sigma model with two hypermultiplets of a vector
charge $(1,-1)$. The D-flatness conditions of this model read as
 \bea (
|\phi^1_1|^2-|\phi^1_2|^2) -( |\phi^2_1|^2-|\phi^2_2|^2)&= &\xi^3
\\
 \phi^1_1
\overline{\phi}_{1}^2-\phi_2^1 \overline{\phi}_{2}^2&=&0
\\
\phi_1^2 \overline{\phi}^{1}_1-\phi^2_2 \overline{\phi}^{1}_2&=&0.
\eea
 Permuting the role of $\phi^1_2$ and $\bar \phi^2_2$,  and
making the following field change $\varphi_1=\phi^1_1$
$\varphi_2=-\bar \phi^2_2$ $\psi_1=\phi^2_1$ and $\psi_2=\bar
\phi^1_2$, the constraint equations (2.12-14) become \bea
 ( |\varphi_1|^2+|\varphi_2|^2) -( |\psi_1|^2+|\psi_2|^2)&=
&\xi^3 \\
 \varphi_1
\overline{\psi}_1+\varphi_2 \overline{\psi}_{2}&=&0
\\
\overline{\varphi_1} {\psi}_1+\overline{\varphi_2} {\psi}_2&=&0
 \eea
 and describe a cotangent bundle over  a $\bf CP^1$ projective.
 In this way, the  $\bf CP^1$ is defined by the following equation:
 \be
  |\varphi_1|^2+|\varphi_2|^2=\xi^3. \ee
     Recall, in passing, that the cotangent bundle over  $\bf CP^1$,
     which is known by the resolved $A_1$
    singularity of K3 surfaces,  is
 isomorphic to  $C^2 \over Z_2$ and plays a crucial role in the
 study of the  non perturbative limit of type II superstring dynamics in
  six and four  dimensions
  [13,14,15].
 The second example we want
to  consider  deals with the generalization of the first one. This
concerns
 a $2D$ $N=4$
$U(1)^r$ linear sigma model with $(r+1)$  hypermultiplets of  a
matrix charge  satisfying (1.1).  Using the same procedure, the
 D-flatness conditions (2.1) become: \bea
 ( |\varphi_{a-1}|^2+| \varphi_a|^2) -( |\psi_{a-1}|^2+|\psi_{a}|^2)&=
&\xi^3_a \\
 \psi_{a-1}
\overline{\varphi}_{a-1}+\varphi_a \overline{\psi}_{a}&=&0
\\
\varphi_{a-1} \overline{\psi}_{a-1}+{\psi}_{a}
\overline{\varphi}_a&=&0. \eea  The solution of these equations
describes the cotangent bundle over $r$ intersecting  complex
curves $\bf CP^1$.  In the limit when all $\xi^3_a$ go to zero, the
$CP^1$'s shrink  and one ends with the $A_r$ singularity of local
K3 surfaces. Note that this example has been used in [11] to construct
seven real dimensional manifolds $ X_7$ of $G_2$ holonomy  group
from the quotient of $X_8$ hyper-Kahler eight real dimensional
manifolds by an $U(1)$ group symmetry. These eight dimensional spaces are
obtained using $2D$ $N=4$ $U(1)^{(r-1)}$ linear sigma model with
$(r+1)$ hypermultiplets, where the missing $U(1)$ invariance is
explored to get  the quotient $X_8\over U(1)$ of  $G_2$ holonomy
group [11]. In what follows we want to give a new class  of $ X_8$
manifolds, which will be called toric-hyper-Kahler Calabi-Yau
fourfolds $(CY^4=X_8)$ by introducing the $ADE$ Cartan matrices
instead of the gauge matrix charge  given in   equation (1.1).
 \section{ Toric-Hyper-Kahler eight manifolds with
 Calabi-Yau condition}
 We start this section by recalling that complex Calabi-Yau
 manifolds are the  best ingredients for obtaining  semi-realistic
 models of superstrings/M/F-theory [18,19,20], with minimal supercharges in lower
 dimensions. In particular, for later use,  Calabi-Yau fourfolds,
 compact, non compact, singular or non-singular, are  considered as
  ways  for getting
 $N=1$ supersymmetric models  in four dimensions  from the F-theory
  compactifications [20,21]. In M-theory  context, compatifications on manifolds of  $G_2$
    holonomy    can be effectively described by four dimensional $N=1$
    supersymmetry. Furthermore, from  supersymmetry breaking
    viewpoint, the above  geometries, which   preserve both the same
    supercharges  in particular $1\over 8$ of initial ones
 of the   uncompactified theory, have a similar role in  superstrings and M-theory
 compactifications.  From this physical argument and  the string duality
 results,
 connecting type IIA and type IIB strings, we think that there
 are, at least,  two natural questions. The latters are as
 follows:
 (1) Exist  there a  four dimensional duality connecting M-theory
 on manifolds of $G_2$ holonomy and F-theory on  Calabi-Yau
  fourfolds?. (2) Or exist there a link between the corresponding
  geometries,(manifolds of $G_2$ holonomy  and  Calabi-Yau
  fourfolds)?. These  questions, which  are quite similar to the
  link  between  M-theory on manifolds of $ G_2$  holonomy
  and heterotic strings on Calabi-Yau threefolds, need deeper
  study. However, here we try to give a modest comment on the the second one; while
   the first  one  will be dealt with in   future work.  This
    comment is based on the following  known points:\\
  (i) Manifolds  with $ G_2$  holonomy  can be constructed as $
  U(1)$ quotients of eight manifolds.\\
  (ii) The maximal group of automorphisms in eight dimensions is
  $SO(8)$. Using Dynkin geometries,  this group, including  the SU(4) group,  can give  the
   $G_2$ group.
  \\
  (iii)  Eight manifolds can have hyper-Kahler constructions in
  terms of $ N=4$ sigma model.\\
  Combining these points with the Calabi-Yau  condtuion, $ \sum_i
  q^a_i=0$, in sigma model approach, one may  say that   seven real dimensional manifolds
   of $G_2$
 holonomy  group  may be  constructed from  hyper-Kahler eight manifolds with
 the
 Calabi-Yau condition. In what follows we refer to such manifolds
  as Calabi-Yau fourfords geometries. In this way, the $G_2$ manifolds     can be obtained
using  quotients
  by  one  finite
 circle,
 preserving  the
 supercharges.
 In this present study, using  similar ideas  of [11], we would like to
   discuss the  construction of   seven dimensional
 manifolds  with $G_2$
  holonomy group  from Calabi-Yau
  fourfolds
 geometry physics data,  but with  a different realization of the $
 U(1)$ group symmetry
  for
 obtaining  the quotient.  This study   involves two steps.
First we  will  introduce, in the field theoretical construction
of Calabi-Yau fourfolds $X_8$,  the ADE Cartan matrices as $2D$
$N=4$
 linear sigma model matrix gauge charges. Second, mimicking the method  of
[11] and  using toric geometry circle actions, we will discuss
quotients  $X_8\over U(1)$ of  $G_2$ holonomy group which  will be
given in the next section. Roughly speaking the toric-hyper-
Calabi-Yau
 fourfolds $CY^4=X_8$ may be viewed
 as the moduli space of $2D$ $N=4$
 supersymmetric $U(1)^r$ gauge theory with $(r+2)$ $\phi_i^{\alpha}$
  hyper-multiplets $(4(r+2-r)=8)$ with
 a matrix charge $q_i^a$ with the Calabi-Yau condition (1.2).  In what  follows, we
  will consider a matrix
 charge going beyond the equation (1.1). Our choice will be given
 by ADE  Cartan matrices. For simplicity, we  first consider the $A_{r}$ Lie algebra
   where
  the
  Cartan matrix  is given by
 \begin{equation}
\ q^a_i=-2\delta^a_i+\delta^a_{i-1}+\delta^a_{i+1},\quad
a=1,\ldots,r,
\end{equation}
 satisfying  naturally the Calabi-Yau condition
 $\sum_iq_i^a=0$. Putting  these   equations into the D-flatness equations(2.1), one
gets the following  system of $3r$ equations:
 \bea
 (|\phi^1_{a-1}|^2+|\phi^1_{a+1}|^2-2|\phi^1_{a}|^2)-(|\phi^2_{a-1}|
 ^2+|\phi^2_{a+1}|^2-
 2|\phi^2_a|^2)&=&\xi_a \\
\phi^1_{a-1}\overline{\phi^2}_{a-1}+\phi^1_{a+1}\overline{\phi^2}
_{a+1}-2\phi^1_{a}
\overline{\phi^2}_{a}&=&0
\\
\phi^2_{a-1}\overline{\phi^1}_{a-1}+\phi^2_{a+1}\overline{\phi^1}
_{a+1}-2\phi^2_{a} \overline{\phi^1}_{a}&=&0 . \eea We first solve
these equations for   the simple example of $U(1)$ gauge theory.
Then we  will give the result for  the $U(1)^r$ gauge model.
 For
$r=1$,  the above equations reduce  to
 \bea
 (|\phi^1_{0}|^2+|\phi^1_{2}|^2-2|\phi^1_{1}|^2)-(|\phi^2_{0}|
 ^2+|\phi^2_{2}|^2-
 2|\phi^2_1|^2)&=&\xi \\
\phi^1_{0}\overline{\phi^2}_{0}+\phi^1_{2}\overline{\phi^2}
_{2}-2\phi^1_{1} \overline{\phi^2}_{1}&=&0
\\
\phi^2_{0}\overline{\phi^1}_{0}+\phi^2_{2}\overline{\phi^1}
_{2}-2\phi^2_{1} \overline{\phi^1}_{1}&=&0.\eea
 To handle these  D-terms equations, it should  be interesting  to note that they are
  quite similar to
 equations (2.10-11), and also (2.12-14).  After permuting the role of $\phi^1_2$ and
 $\overline{\phi^2}_{2}$, equations may be rewritten as
 \bea
 (|\phi^1_{0}|^2+|\phi^1_{2}|^2+2|{\overline {-\phi^2_{1}}}|^2)-(|\phi^2_{0}|
 ^2+|\phi^2_{1}|^2+
 2|\phi^1_1|^2)&=&\xi \\
\phi^1_{0}\overline{\phi^2}_{0}+\phi^1_{2}\overline{\phi^2} _{2}+2
\phi^1_{1} \overline{{(-\phi^2}_{1})}&=&0
\\
\phi^2_{0}\overline{\phi^1}_{0}+\phi^2_{2}\overline{\phi^1}
_{2}+2(-\phi^2_{1}) \overline{\phi^1}_{1}&=&0.\eea  Making the
following  field changes  \bea \phi^1_{0}=\varphi_1,\quad
\phi^2_{0}=\psi_1 \nn\\ \phi^1_{1}=\varphi_2,\quad
\phi^2_{1}=\psi_2 \nn\\ - {\overline \phi^2_{1}}=\varphi_3,\quad
{\overline \phi^1_{1}}=\psi_1, \nn
 \eea the above equations
become
 \bea
 (|\varphi_{1}|^2+|\varphi_{3}|^2+2|\varphi_{2}|^2)-(|\psi_{1}
 |^2+|\psi_{3}|
 ^2+2|\psi_2|^2)
 &=&\xi^3 \\
\varphi_{1}\overline {\psi_{1}}+\varphi_{3}\overline
{\psi_{3}}+2\varphi_{2}\overline {\psi_{2}} &=&0
\\\overline{ \varphi_{1}} {\psi_{1}}+\overline
{\varphi_{3}}\psi_{3}+2\overline {\varphi_{2}}\psi_{2}&=&0. \eea
 Using similar analysis of the previous section, one sees that the above equations
  describe a cotangent bunble over  $\bf WCP^2_{1,2,1}$ weighted  projective
  space. A way to to see this feature is  to  use the link between  $N=2$ sigma model and
  toric geometry
 technics.  Indeed,   taking  $\psi_{1}=\psi_{2}=\psi_3=0$,  equations (3.11-13)
 reduce to
\be |\varphi_{1}|^2+|\varphi
_{3}|^2+2|\varphi_{2}|^2
 =\xi^3
 \ee
which   can be encoded in a toric diagram. In this diagram,  one
has three vectors $v_1$, $v_2$ and $ v_3$ in $Z^2$ lattice  such
that \be v_1+v_3+2v_2=0 \ee
 where the  coefficients  of $v_i$  are exactly  the ones of  $|\varphi_i|^2$ in
 (3.14). Note that equation (3.15) describes a particular geometry of the one given in (2.4).
  Using the  toric geometry language,
   equation (3.15)   defines  naturally  a  $\bf WCP^2_{1,2,1}$ weighted  projective
  space,  where  $\xi^3$
is a Kahler real parameter controlling its  seize.   The equation
(3.11-13), for generic value of $\psi_{i}$, can be interpreted to
mean that  $\psi_{i}$  parameterizes  the fiber   cotangent
directions over $\bf WCP^2_{1,2,1}$.  Since the subset  of (3.11)
with $\psi_{i}=0$ is a  $\bf WCP^2_{1,2,1}$ weighted  projective
space and $\varphi_{1}\overline {\psi_{1}}+\varphi_{3}\overline
{\psi_{3}}+2\varphi_{2}\overline {\psi_{2}} =0$ is the analogue of
equation (2.11), thus the the space of solutions of (3.11-13) is
isomorphic to the cotangent space over $\bf WCP^2_{1,2,1}$, $
T^*(\bf WCP^2_{1,2,1})$. In the general case corresponding  to the
$U(1)^r$ gauge theory, if we take the all  $\xi_a$'s are  no zero,
it not too difficult to see that
 equations (3.2-4) describe the cotangent bundle over  $r$ intersecting
  $\bf WCP^2_{1,2,1}$ weighted
 projective spaces.  This means that the base geometry, of the cotangent  bundle,
  consists of
 $r$ intersecting
  $\bf WCP^2_{1,2,1}$ according to the $A_r$ Dynkin  diagram, instead of  one
  projective space in the  case of  $U(1)$ gauge theory. In the limit
   that some $\xi_a$'s
   are   zero, we obtain a singular geometry. Actually,
   this geometry may be used to extend the
 intersecting  $\bf CP^1$
  projective spaces of ALE spaces involved  in
 the geometric engineering method of the quantum  field theory
 [13,14,15]. We will conclude this section by  noting that this  analysis
   of  the  $A_r$ Lie algebra may be  extended
 to  the others DE Lie algebras. However, these  algebras contain trivalent
 vertex Dynkin  geometries which complicates  the  computation.
 Recall that  the trivalent Dynkin geometry involves a central node
 intersecting three other nodes  once; moreover  this  geometry  has been  used in the
  geometric engineering of quantum  field theories, in particular  in the introduction
   of fundamental matters in a chain  of $SU$ product gauge group with $N=2$
  bifundamental matters.  In toric sigma  model
 approach, the corresponding  vector charge,  up  the Calabi-Yau
 condition, is given  by
 $$ q_i=(0,\ldots, -2,1,1,1,0,\ldots,0,-1),$$ instead of the bivalent  geometry (3.1).  A priori there  are
  different ways one may follow  to overcome this  problem. A naive way  is to    delete these
   trivalent  vertices. In this  case,   the D-flatness constraint
 equations have similar solutions of the  $A_n$ Lie algebra.  However a tricky method is
 to leave  and use  the trivalent geometry  results involved   in  the  elliptic fibrations
 singularities over the complex  plane.  In this way,   the base geometry  of $X_8$ may be
  given by three chains of  intersecting  $\bf WCP^2_{1,2,1}$ according to
the trivalent geometry. \\
In what follows we would like to  discuss
 the  corresponding  seven  real dimensional  manifolds  of $G_2$ holonomy group using  $U(1)$
  quotients.
 Similarly to the ideas of [11], we should look for a $U(1)$  group symmetry acting
  on  $X_8$. As mentioned before,  there are
   many ways one may  follow to  choosing the $U(1)$ group action  of
   $X_8$.  In this regard, the solutions differ by  what kind of $U(1)$
   symmetry is chosen. Two kinds of  solutions  are given in [11].
    But here we
  will  consider another way. The latter is inspired from the toric
   geometry circle actions.
  \section{On the quotient space $X_7= {X_8\over U(1)}$ of $G_2$ holonomy }
  Having constructed  toric-hyper-Kahler Calabi-Yau fourfolds $X_8$
  associated to
   ADE Cartan matrices sigma
  model gauge  charges, we are now in the position
   to carry out
   quotient spaces $X_7= {X_8\over U(1)}$  of  $
   G_2$ holonomy group using circle actions involved  in toric varieties.
 Before doing this, let us tell some
    things about toric  geometry.  The letter is  a powerful tool for
     studying
        $n$-dimensional complex
       manifolds exhibiting toric circle  actions $U(1)^n$ which allow
         to encode the geometric properties of the complex spaces
          in terms of simple combinatorial data of polytopes
           ${\Delta}$ of the $R^n$ real space [22,23,24,25].  The simple example of
           toric varieties is
the  complex plane $\bf C$.  The latter admits a  $U(1)$  toric action
\be
z\to ze^{i\theta}, \ee which has a fixed point at $z=0$. Thus the
toric geometry of  $\bf C$     can be viewed as a circle fibered
on a half line parameterized  by $|z|$. The circle,  which
determined by the  action of $\theta$,  shrinks at $z=0$. This
realization can be  generalized easily to $\bf C^n$ space   where
we have a $T^n$ fibration,  parameterized by the angular
coordinates $\theta_i$, over  a  $n$-dimensional real base
parameterized  by $|z_i^2|$. The second  example we want to give
is the  $\bf CP^1$ projective space.   This space has also  a
$U(1)$ toric action having two fixed  points describing
  respectively  north and south poles of the two sphere $S^2\sim
  \bf CP^1$.     Thus   the toric  geometry of  $\bf CP^1$  is given by an
   interval  fibered
by
   $ S^1$ with zero  size
  at the endpoints of the  interval. Using these ideas,
   the cotangent bundle over  $\bf CP^1$   can be also viewed as a toric space. In this way,
    we have two  circle
     actions on
   this space. The first one is the one  corresponding to the action
   on the $\bf CP^1$ base space  and the   other  circle  acts on the fiber cotangent
    direction. Our next example will be   the two complex dimensional projective space
     $\bf CP^2$. The latter  has   a  $U(1)^2$ toric action exhibiting three fixed
     points defining a triangle in the $\bf R^2$ real space.
   The toric geometry  of this  manifold is
   described by a triangle  of   $\bf R^2$  fibered   by a  two real  dimensional torus $T^2$
    which degenerates  to a $S^1$  circle  on  the three edges and  shrinks
     to a point  on
    the endpoints.  The cotangent  bundle over  $\bf CP^2$ is a
     4-dimensional  (eight real)
     local toric  geometry,
  where  we have two extra circle  actions coming from the fiber
    cotangent directions.  Note  that this analysis is similar for the
    $\bf WCP^2$, in particular   $\bf WCP^2_{(1,2,1)}$, and can be extended
     easily to higher
    dimensional (weighted)  projective spaces. In what follows we
    will consider the above toric geometry circle  actions to identify the
    $U(1)$ group symmetry of  quotient spaces $X_7= {X_8\over
    U(1)}$. \\ Let us consider the simple example of  the  $U(1)$ gauge theory with three
     hypermultiplets.
    In this case  the   geometry  $X_8$ can be viewed  as $\bf C^2$ bundle over  a
      $\bf WCP^2_{(1,2,1)}$. This manifold  has four toric geometry  circle actions
      $U(1)^2_{base}\times U(1)^2_{fiber}$;  two of them   correspond to the
     $\bf WCP^2_{(1,2,1)}$'s
    base space
     denoted by  $U(1)^2_{base}$
    while  the remaining ones $U(1)^2_{fiber}$ act on  the  fiber
    cotangent directions. In what follows,  we want to divide by one  finite circle toric geometry
     action for  obtaining   seven real manifolds.  Mimicking the analysis of
     [11] and
     identifying    the $U(1)$ group symmetry of the quotient  with one finite
      fiber circle action
\be
X_7={X_8\over U(1)_{fiber}},
 \ee
    we  can obtain a 7-dimensional geometry. Since ${{\bf C^2}\over U(1)}={\bf R}^+\times
     {\bf C}$,  the quotient  space  is now a ${\bf R}^+\times
     {\bf C}$ bundle over  a  $\bf WCP^2_{(1,2,1)}$. By compactifying the
 ${\bf C}$ complex plane, which can be done by adding a point at infinity,
 this  space will be  a  ${\bf R}^+\t {S^2}$ bundle over  $\bf WCP^2
      _{(1,2,1)}$.  Similarly to [11], this geometry is  a  cone on  a  $ {S^2}$ bundle over
        $\bf WCP^2
      _{(1,2,1)}$ of $G_2$ holonomy.   More generally,   if  we consider the
         $U(1)^r$ gauge theory  with  the $ A_r$ Cartan matrix  gauge charges and
          $(r+2)$ hypermultiplets, then   the quotient space is  a cone on  a  $ {S^2}$ bundle over
     $r$ intersecting   $\bf WCP^2_{(1,2,1)}$ weighted  projective spaces  according to
      the $A_r$ Dynkin diagram.\\
    Finally,  a naive way to study the singularities of these  $X_7$ manifolds
    is to consider the identification structure of the
    weighted projective spaces. The latters are not generally
    smooth because non trivial fixed points under the variable
    identifications lead to singularities. To see this feature,
    consider the identification structure of  $\bf WCP^2_{1,2,1}$  defined by
    introducing three homogeneous complex coordinates $z_1,z_2,z_3$
    not all of them
    simultaneously zero with a projective relation:
    \be
    (z_1,z_2,z_3)\equiv(\lambda z_1,\lambda^2 z_2,\lambda z_3).
    \ee
Note, in passing,  that these  $(z_1,z_2,z_3)$ homogeneous complex
coordinates can be related  respectively to  $\psi_1, \psi_2$  and
$\psi_3$  fields  of the  sigma model
 construction.  Finally, it is not hard to show that this space
is singular. Indeed, if  we take $\lambda=-1$, equation (4.3)
reduces to
\be
    (z_1,z_2,z_3)\equiv(-z_1, z_2,- z_3),
    \ee
and so we have a $Z_2$ orbifold singularity at $(z_1, z_2,
z_3)=(0,z_2,0)$.

    \section{ Discussion and Conclusion}
 In this paper, we have contributed in  the  M-theory compactifications
 to four dimensions. This involves the compactification on seven
 manifolds  of  $G_2$ holonomy  group, leading to $N=1$  four
 dimensional
 supersymmetric models. In particular, we have constructed a new
 class of toric-hyper-Kahler eight manifolds giving $ G_2$ holonomy  spaces after
 dividing by   one finite  toric geometry circle action. This building has been proceeded in
 two steps. We  have first introduced the ADE Cartan matrices as matrix
 gauge charges in  the  $N=4$ $2D$ field theoretical construction of
  toric-hyper-Kahler eight manifolds $X_8$. In particular, the solution
 for the  $A_r$ Lie algebra is described by the cotangent bundle over
$r$ intersecting $\bf WCP^2_{1,2,1}$ weighted projective spaces
according to the $A_r$ Dynkin diagram. Actually
 these spaces may  extend the  geometry of
 $A_r$ ALE space, described  by $2D$ $N=2$ type IIA superstring sigma model
  used in the geometric engineering method. Second we have used the toric
  geometry circle
  actions of $X_8$ to build  quotients  $X_7= {X_8\over
 U(1)}$ of  $G_2$ holonomy group.\\
\\ \\ {\bf Acknowledgments}\\
    Adil Belhaj would like to thank E.H. Saidi for  valuable
    discussions. He is grateful to U. Lindstrom for sending him  the paper
    [17].
    This work is supported by SARS,
    programme de
soutien \`a la recherche scientifique; Universit\'e Mohammed
V-Agdal, Rabat.\\

 \end{document}